\documentclass[letterpaper]{aa}
\usepackage{txfonts}
\usepackage{graphicx}
\usepackage{natbib}
\bibpunct{(}{)}{;}{a}{}{,}

\begin{document}

\title{The effect of type I migration on the formation of terrestrial planets in hot-Jupiter systems}
\author{Martyn J. Fogg \& Richard P. Nelson.}
\institute{Astronomy Unit, Queen Mary, University of London, Mile
End Road, London E1 4NS.\\
\email{M.J.Fogg@qmul.ac.uk, R.P.Nelson@qmul.ac.uk}}
\date{Received/Accepted}

\abstract
{Our previous models of a giant planet migrating through an inner
protoplanet/planetesimal disk find that the giant shepherds a
portion of the material it encounters into interior orbits, whilst
scattering the rest into external orbits. Scattering tends to
dominate, leaving behind abundant material that can accrete into
terrestrial planets.}
{We add to the possible realism of our model by simulating type I
migration forces which cause an inward drift, and
strong eccentricity and inclination damping of protoplanetary bodies.
This extra dissipation might be expected to enhance shepherding at the expense
of scattering, possibly modifying our previous conclusions. }
{We employ an N-body code that is linked to a viscous gas disk
algorithm capable of simulating: gas accretion onto the central
star; gap formation in the vicinity of the giant planet; type II
migration of the giant planet; type I migration of protoplanets; and
the effect of gas drag on planetesimals. We use the code to re-run
three scenarios from a previous work where type I migration was not
included.}
{The additional dissipation introduced by type I migration enhances
the inward shepherding of material but does not severely reduce
scattering. We find that $>$ 50\% of the solids disk material still
survives the migration in scattered exterior orbits: most of it well
placed to complete terrestrial planet formation at $<$ 3~AU. The
shepherded portion of the disk accretes into hot-Earths,
which survive in interior orbits for the duration of our simulations. }
{Water-rich terrestrial planets can form in the habitable zones of
hot-Jupiter systems and hot-Earths and hot-Neptunes may also be
present. These systems should be targets of future planet search
missions.}

\titlerunning{Effect of type I migration on formation of terrestrial planets in hot-Jupiter systems}
\authorrunning{M.J. Fogg \& R.P. Nelson}

\maketitle

\section{Introduction.}\label{intro}

The core accretion hypothesis of giant planet formation postulates
that giant planets originate in a cool location in a protoplanetary
disk, beyond the snowline where the condensation of ices boosts the
mass of solid material present \citep[e.g.][]{pollack,papaloizou2}.
This, and the fact that remoteness from the central star enhances
the gravitational reach of accretion, may allow a
$\sim10~\mathrm{M}_\oplus$ core to grow and then accumulate a
massive gaseous envelope before the loss of the nebular gas which is
observed to occur $\sim$~1--10~Myr after star formation
\citep{haisch}. The alternative gravitational instability model also
predicts that giant planets will form in the cool outer disk regions
\citep{boss}. Giant planets are also expected to migrate inward from
their original formation location due to gravitational interactions
with the disk, either during the core stage via type I migration,
where the body remains embedded in the gas
\citep[e.g.][]{ward2,papaloizou1,tanaka1,tanaka2}, or via type II
migration when the planet has grown massive enough ($\gtrsim
100~\mathrm{M}_\oplus$) to open a gap in the gas and becomes coupled
to the inward viscous evolution of the disk
\citep[e.g.][]{lin1,lin2,ward2,nelson1}. According to recent models
of the late stages of solar system formation, which appear to
explain the current architecture of the solar system
\citep{tsiganis,morbidelli,gomes}, Jupiter and Saturn appear to have
undergone only limited migration, suggesting that they may have
formed quite late with respect to dispersal of the gas. Their
influence on interior terrestrial planet formation at distances
$\lesssim 2$~AU was probably modest, although recent models suggest
secular interactions may have played an important role during gas
disk dispersal \citep{nagasawa}. In contrast, many giant exoplanets
show evidence for substantial migration, being found well within
what would have been their parent nebula's snowline \citep{butler},
the most extreme examples being known as hot-Jupiters with examples
found at distances $\lesssim 0.1$~AU down to just a few stellar
radii. These planets must have passed through their inner systems at
an early time ($\sim$~1--10~Myr) and would inevitably have had a
strong influence on the formation of rocky planets that would
otherwise be expected to grow there on timescales of
$\sim$~10--100~Myr \citep{chambers2,kleine,halliday,obrien,nimmo}.

It is expected that terrestrial planets external to a hot-Jupiter
would have stable orbits \citep{jones} and should be able to form
from any available pre-planetary material with a period ratio
$\gtrsim$~3 \citep{raymond1}. However, until recently it has been a
common approach in both astrophysical and astrobiological literature
to assume that terrestrial planets could not have formed in
hot-Jupiter systems due to the disruptive effect of the giant
planet's migration which is deemed to have cleared the inner system
of planet-forming material
\citep[e.g.][]{ward1,lineweaver1,armitage1,lineweaver2}. However,
recent work by two groups who have modeled terrestrial planet
formation in the presence of type II giant planet migration have
shown that this assumed clearing does not happen
\citep{fogg1,fogg2,raymond2,fogg3,mandell}. Their simulations show
that the giant planet shepherds the solids disk inward, compacting
it and exciting the orbits of objects caught at first order
resonances. Much of this excited material eventually experiences a
close encounter with the giant planet and is expelled into an
exterior orbit. The net effect of the giant's passage is therefore
not the destruction and disappearance of the inner system disk, but
instead a modest dilution and strong excitation and radial mixing of
solid material. Over the entire course of the migration, a new disk
of solid material builds up in orbits external to the final position
of the hot-Jupiter where terrestrial planet formation can resume.
The results of further simulation of accretion in this scattered
disk suggest that terrestrial planets should be commonplace in
hot-Jupiter systems: worlds that might possibly be water-rich due to
strong inward mixing of material from beyond the nebular snowline
\citep{raymond2,fogg3,mandell}.

All these papers agree that the outcome of giant planet migration
through an inner system solids disk is to partition most of its mass
into the exterior scattered remnant described above and an interior
remnant resulting from shepherded material that is not scattered by
the giant. This shepherded fraction is compacted into the restricted
volume close to the central star, where dynamical times are short,
resulting in a rapid burst of accretion with the potential to form
stable hot-Neptune or hot-Earth type planets orbiting interior to
the final position of the giant \citep[see also][]{zhou}. Comparing
results however reveals that the relative predominance of scattering
versus shepherding is significantly influenced by the strength of
dissipative forces (gas drag and dynamical friction) operating in
each particular model. Whilst interior and exterior remnants are
found in all cases, a distinct trend is evident favoring shepherding
under conditions of strong dissipation and scattering when
dissipation is weaker. For example, the model in \citet{fogg1}
assumed a steady-state gas disk with a fixed $r^{-1.5}$ surface
density profile which maximized both the influence of gas drag on
planetesimal sized objects and hence their ability to damp the
orbits of protoplanets via dynamical friction. The formation of
shepherded hot-Neptunes and hot-Earths was a common outcome of these
simulations with as little as $\sim$26\% of the original disk
material being scattered into the external disk in the youngest
scenarios. (In the case of constant gas density the average strength
of dynamical friction still falls with disk maturity as
planetesimals are cleared through accretion.) In contrast, in their
latest model \citep{fogg3} which incorporates a time dependent,
viscously evolving, gas disk with a partial central cavity and
annular gap centered on the giant planet's orbit, dissipation is
much lower and declines with time, especially in the inner regions
of the system. In this perhaps more realistic case, hot-Earths start
to form in the shepherded portion of the disk but are almost always
scattered or accreted by the giant at late times when it has
migrated down to $a \lesssim 0.5$~AU. Irrespective of the maturity
of the solids disk at the start of migration, $>60\%$ of its mass is
always scattered into an external disk. Dissipation levels within
the model of \citet{raymond2} and \citet{mandell}, with its linearly
declining, fixed surface density profile gas disk, appear to lie
intermediate between the two models of Fogg \& Nelson. Shepherded
hot-Earths sometimes survive at the end of their simulations and
sometimes not, although a rigorous comparison between their work and
that of Fogg \& Nelson is not possible due to differences in assumed
initial conditions, and the fact that in most of their simulations
the giant planet comes to rest at a radial distance of $>0.1$~AU.

One potentially important physical process neglected in these models
is type I migration which operates on sub-gap opening bodies of
$\sim0.1-100~\mathrm{M}_\oplus$
\citep[e.g.][]{ward2,papaloizou1,tanaka1,tanaka2}. Type I migration
is generated by the asymmetric torques exerted on a protoplanet from
the wakes it creates in the gas disk and is expected to impart an
inward radial drift and strong eccentricity and inclination damping
that increases in effectiveness linearly with protoplanetary mass.
The realism of type I migration has long been controversial as
theoretically predicted spiral-in times can be so short as to
threaten protoplanetary survival and to render an explanation of the
solar system's architecture particularly difficult. Many planet
formation models therefore ignore this process entirely. It might be
that type I migration does not operate as rapidly as predicted: one
proposal being that stochastic torques from density fluctuations in
a turbulent disk could superimpose a random walk in semi-major axis
over the smooth inward drift predicted by theory
\citep{nelson2,laughlin,nelson3}. However, it is becoming apparent
that even near-nominal rates of type I migration are not necessarily
fatal to planet formation and survival. Work by \citet{mcneil} has
shown that it is possible to grow a terrestrial planetary system in
a simulation including type I migration forces by enhancing the
original quantity of solid material present whilst including a
rapidly dissipating gas disk. The model of \citet{daisaka} succeeds
in retaining sufficient material in the terrestrial region by
invoking a shorter gas depletion timescale in the inner system than
the observational value (based on infrared observations at
$a\sim100$~AU), or if the the initial gas to dust ratio is smaller
than the conventional minimum mass solar nebula (MMSN) model. The
problem with preventing the more massive giant planet cores from
being lost to the central star is of greater severity, but recent
work by \citet{thommes2} suggests that this may be possible late in
the lifetime of the gas disk when accretion and migration timescales
become comparable. Proposed system properties that might widen such
a window for core survival include small planetesimal sizes, low
midplane gas disk viscosities and enhanced collision cross sections
due to core atmospheres \citep{chambers3,thommes3}.

The possible realism of strong type I migration forces is of obvious
and perhaps critical relevance to the above-cited models of giant
planet migration through an inner system solids disk. Even if the
issue of pre-existing type I migration of the giant planet's core is
neglected and the assumption of the fully formed giant planet
undergoing type II migration is retained, type I migration forces
would still be expected to operate on the protoplanetary components
of the inner system disk, introducing another source of damping
additional to dynamical friction. Extra dissipation would be
expected to affect the partitioning of inner system material by
enhancing the shepherded fraction at the expense of the scattered
fraction, potentially invalidating all the previous claims that
sufficient solid material survives the passage of the giant to
permit the formation of terrestrial planets. Our primary motivation
is to investigate this issue here by adding type I migration forces
to the model of \citet{fogg3} and re-running one early and two late
scenarios from that paper. We note that additional sources of
damping may also be important such as damping by collisional debris,
but inclusion of these effects is beyond the scope of our model. In
Section 2 we outline the additions to our model and the initial
conditions of the simulations; in Section 3 the results are
presented, discussed, and compared to previous work; in Section 4 we
consider some caveats, and in Section 5 we offer our conclusions.

\section{Description of the model.}\label{description}

We model planetary accretion using a modified version of the
\emph{Mercury 6} hybrid-symplectic integrator \citep{chambers1}, run
as an $N + N'$ simulation, where we have $N$ protoplanets embedded
in a disk of $N'$ ``super-planetesimals" -- particles with masses a
tenth that of the initial masses of protoplanets which act as
statistical tracers of a much larger number of real planetesimals
and are capable of exerting dynamical friction on larger bodies
\citep[e.g.][]{thommes1}. The protoplanets (and the giant when it is
introduced) interact gravitationally and can accrete and merge
inelastically with all the other bodies in the simulation, whereas
the super-planetesimal population is non-self-interacting.
Super-planetesimals however are subject to a drag force from their
motion relative to the nebular gas that is equivalent in its
dynamical action to the gas drag that would be experienced by a
single 10~km radius planetesimal. A detailed outline of these
aspects of our model is given in \citet{fogg1}.

We model the evolution of the nebular gas using a 1-D viscous disk
model that is linked to the N-body code with a time-step
synchronization routine. This model solves numerically a viscous gas
disk diffusion equation that is modified to include the tidal
torques exerted by an embedded giant planet \citep{lin1,takeuchi}
and its implementation is fully described in \citet{fogg3}. The gas
responds under the influence of this algorithm by depleting over
time via viscous accretion onto the central star; by opening up an
annular gap centered on the giant planet's orbit; and by forming a
partial inner cavity due to dissipation of propagating spiral waves
excited by the giant planet. The back reaction of these effects on
the giant planet is resolved as torques which drive type II
migration in a fully self-consistent manner.

\subsection{Type I migration forces.}\label{typeI}

We have added type I migration forces to our model, which are
applied to protoplanets only, by implementing the approach of
\citet{cresswell} who adopted the migration time prescription of
\citet{tanaka1} and eccentricity damping time prescription of
\citet{tanaka2} and modified them with factors derived by
\citet{papaloizou1} to describe evolution in the case of large
eccentricity.

Their formula for the type I migration time is:

\begin{equation}\label{migtime}
t_\mathrm{m} =
\frac{2}{2.7+1.1\beta}\left(\frac{M_*}{m}\right)\left(\frac{M_*}{\Sigma_\mathrm{g}
a^2}\right)\left(\frac{h}{r}\right)^2\left(\frac{1+\left(\frac{e~
r}{1.3 h}\right)^5}{1-\left(\frac{e~r}{1.1 h}\right)^4}\right)
\Omega^{-1} ,
\end{equation}
where $M_*$ is the mass of the central star, $\Sigma_\mathrm{g}$ is
the gas surface density in the vicinity of the planet, $h$ is the
gas scale height, and $m, a, r, e$ and $\Omega$ are the planet's
mass, semi-major axis, distance from the star, orbital eccentricity
and orbital frequency respectively. The factor $\beta$ in the first
term is the gas disk surface density profile index
$(\Sigma_\mathrm{g} \propto r^{-\beta})$, which we take to be fixed
at its initial value of $\beta = 1.5$, even though $\beta$ falls in
value during the evolution of the inner disk as gas drains onto the
star. In practise, this simplification makes only a $\sim 15\%$
difference to $t_m$ at later times and avoids the need for
additional, time-consuming, measurement of $\beta$ at every
protoplanetary location each gas disk time step. Important
behavioral features to note from Eq.~\ref{migtime} are that type I
migration speeds up with an increase in planetary mass and slows
down with a decrease in gas density or increased eccentricity. When
$e >1.1~h / r$, inward migration halts as $t_m$ becomes negative and
only resumes when eccentricity is damped to lower values. We note
that under certain circumstances, such as there being a surface
density jump, or an optically thick disk, or MHD turbulence type I
migration may be substantially modified or reversed
\citep{masset,paardekooper,papaloizou2,nelson3}. Consideration of
these effects goes beyond the scope of this paper.

The control of $\Sigma_g(r,t)$ by our viscous disk algorithm causes
$t_m(r,t)$ to behave in an interesting way. Since $t_m \propto
\Sigma_g^{-1} a^{-2}(h/r)^2 \Omega^{-1}$ and $\Sigma_\mathrm{g}
\propto r^{-\beta}$, and as we use $h/r =
0.047(r/\mathrm{AU})^{1/4}$ AU, we find $t_m \propto r^{\beta}$ and
$\dot{r} \propto r^{1-\beta}$. At early times and in our outer disk,
$\beta \simeq 1.5$ and $\dot{r} \propto r^{-1/2}$, so migration is
faster for interior objects of a given mass. However, the viscous
draining of gas onto the central star rapidly results in much of the
nebula declining to a shallower profile index of $\beta \simeq 1$,
falling further to $\beta \simeq 0.75$ interior to 1~AU (see
Fig.~\ref{figure:1} in Sect.~\ref{initial}); in this situation
$\dot{r} \propto r^{1/4}$, so relative migration rates are slower
for interior objects.

Cresswell and Nelson's formula for eccentricity damping time is:

\begin{equation}\label{ecctime}
t_\mathrm{e} =
\frac{Q_\mathrm{e}}{0.78}\left(\frac{M_*}{m}\right)\left(\frac{M_*}{\Sigma_\mathrm{g}
a^2}\right)\left(\frac{h}{r}\right)^4\left(1+\frac{1}{4}\left(e\frac{r}{h}\right)^3\right)
\Omega^{-1} ,
\end{equation}
where $Q_\mathrm{e} = 0.1$ was chosen as a normalization factor to
get $t_\mathrm{e}$ into good agreement with values measured from
hydrodynamic simulations. We additionally assume that the timescale
for inclination damping is the same as that for eccentricity
damping, i.e. $t_\mathrm{i} = t_\mathrm{e}$. As with the behavior
for inward migration, eccentricity damping is also stronger with
increasing planetary mass and weaker with decreasing gas density or
higher eccentricity.

Given $t_\mathrm{m}$ and $t_\mathrm{e}$, type I migration forces are
then applied via the following accelerations to each protoplanet:

\begin{equation}\label{mig_a}
\mathbf{a}_\mathrm{m} = - \frac{\mathbf{v}}{t_\mathrm{m}} ,
\end{equation}

\begin{equation}\label{mig_e}
\mathbf{a}_\mathrm{d} = -2
\frac{(\mathbf{v}\cdot\mathbf{r})\mathbf{r}}{r^2 t_\mathrm{e}} -
2\frac{(\mathbf{v}\cdot\mathbf{k})\mathbf{k}}{t_\mathrm{i}} ,
\end{equation}
where $\mathbf{v}$ is the protoplanet's velocity vector and
$\mathbf{k}$ is a unit vector in the vertical direction.

\subsection{Initial conditions and running of the simulations.}\label{initial}

As an initial state, we assume the presence of a protoplanetary disk
about a 1~$\mathrm{M}_\odot$ star conforming in properties and
structure to the minimum mass solar nebula model of \citet{hayashi},
but scaled up by a factor of three. We assign a nominal age of this
disk to be $\sim$ 0.5 Myr (this being the t = 0 start time for the
simulations) by which point the evolution of the solid components of
the inner disk is considered to have reached the oligarchic growth
stage described by \citet{kokubo2}. This inner solids disk extends
initially between 0.4--4.0~AU and has a snowline at 2.7~AU beyond
which its surface density is boosted by a factor of 4.2 to account
for ice condensation. The generation of the N-body components of
this initial solids disk assumes initial protoplanetary masses of
0.025 and 0.1~$\mathrm{M}_\oplus$ interior and exterior to the
snowline respectively, spaced approximately 8 mutual Hill radii
apart, with the remainder of the material inventory consisting of
super-planetesimals with a fixed mass of 10\% of the initial masses
of the local protoplanets. Further details are described in
\citet{fogg1} and relevant data are shown in Table~\ref{table:1}
which gives, for zones interior and exterior to the snowline, values
for the total mass of solid material $M_{\mathrm{solid}}$, the
number and mass of protoplanets $N$ and $m_{\mathrm{proto}}$, and
the number and mass of super-planetesimals $N'$ and
$m_{\mathrm{s-pl}}$. The parameter $f_\mathrm{proto}$, at the foot
of Table~\ref{table:1}, is the mass fraction of the solids disk
contained in protoplanets and we use this here as a rough measure of
the evolution of the disk, taking $f_\mathrm{proto} = 0.5$ to denote
the transition between oligarchic and giant impact growth regimes.
We model the gas component of the disk from 0.025--33~AU with an
initial surface density profile of $\Sigma_\mathrm{g} \propto
r^{-1.5}$ which gives an initial mass of 0.0398~$\mathrm{M}_\odot$
$\cong 42~\mathrm{M_J}$. The disk alpha viscosity is $\alpha =
2\times10^{-3}$ which gives a viscous evolution time at 5~AU
$\simeq$ 120\,000 years and a mass depletion e-folding time $\simeq$
580\,000 years (see \citet{fogg3} for details).

\begin{table}
\caption{Data describing initial solids disk set-up} %
\label{table:1}  %
\centering
\begin{tabular}{c| c c| c}
 \hline\hline %
& Rocky Zone & Icy Zone& Total\\
& 0.4--2.7~AU & 2.7--4.0~AU & 0.4--4.0~AU\\
 \hline
$M_{\mathrm{solid}}$ & $9.99~\mathrm{M}_{\oplus}$ &
$14.8~\mathrm{M}_{\oplus}$ & $24.8~\mathrm{M}_{\oplus}$\\ %
 \hline
$m_{\mathrm{proto}}$ & $0.025~\mathrm{M}_{\oplus}$ &
$0.1~\mathrm{M}_{\oplus}$\\ %
$N$ & 66 & 9 & 75\\ %
 \hline
$m_{\mathrm{s-pl}}$ & $0.0025~\mathrm{M}_{\oplus}$ &
$0.01~\mathrm{M}_{\oplus}$\\ %
$N'$ & 3336 & 1392 & 4278\\
 \hline
$f_{\mathrm{proto}}$ & 0.17 & 0.06 & 0.1\\ %
 \hline\hline
\end{tabular}
\end{table}

From $t = 0$ we start by running this combined N-body and viscous
disk model in the absence of the giant planet in order to mature it
to different ages. This is done with a symplectic time step of 8
days and a simulation inner edge of 0.1~AU, interior to which any
solid material is eliminated. Our previous simulations have shown
the importance of this maturation as the relative strengths of
dissipative forces (gas drag, dynamical friction, and the newly
included type I migration) all vary with time, as the gas density
falls and protoplanets grow at the expense of planetesimals. We
generate one ``early" and two ``late" scenarios by ageing the system
for 0.1, 1.0 and 1.5~Myr that are the equivalent of Scenarios I, IV
and V from \citet{fogg3} and to facilitate comparison with this
previous work we keep this nomenclature here.

\begin{figure}
 \resizebox{\hsize}{!}{\includegraphics{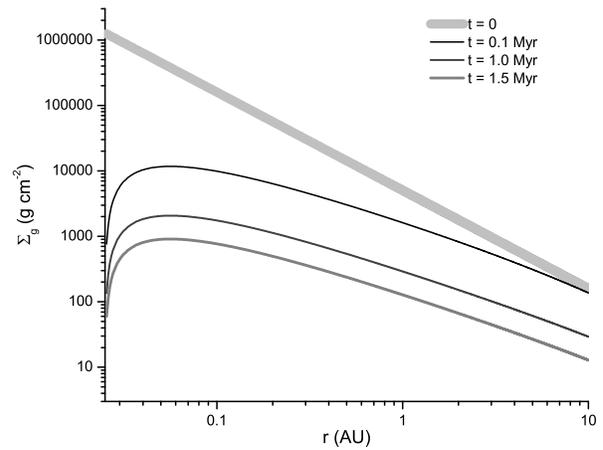}}
 \caption{Evolution of the gas surface density within the inner 10~AU
  of our simulated disk. The upper line is the initial
  $r^{-1.5}~\Sigma_\mathrm{g}$-profile for a $3\times\mbox{MMSN}$ disk. The
  lower curves, in descending order, are the profiles at 0.1,
  1.0 and 1.5~Myr respectively.}
 \label{figure:1}
\end{figure}

\begin{figure}
 \resizebox{\hsize}{!}{\includegraphics{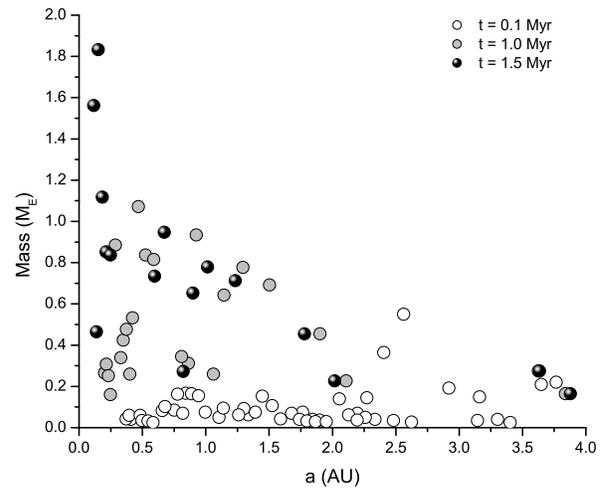}}
 \caption{Protoplanetary masses vs. semi-major axis obtained after evolution
 to the starting points of Scenarios I, IV and V. Substantial inward drift
 is visible at late times due to type I migration.}
 \label{figure:2}
\end{figure}

The initial gas surface density and its profile at these three ages
are plotted in Fig.~\ref{figure:1} and numerical data for the
evolved disks, including the remaining mass of gas
$(M_{\mathrm{gas}})$ and solid material $(M_{\mathrm{solid}})$, the
maximum protoplanetary mass $(m_{\mathrm{max}})$, the number of
surviving particles $(N~\&~N')$, and the protoplanet mass fraction
$(f_\mathrm{proto})$ are given in Table~\ref{table:2}. Large
reductions in gas density are illustrated, especially in the inner
regions of the system. Particle numbers fall and $m_{\mathrm{max}}$
and $f_\mathrm{proto}$ rise with time as expected and by the time of
the latest scenario at $t = 1.5$~Myr, some solid material has been
lost within 0.1~AU. It might be thought that this loss (just $\sim
5\%$ of our solids inventory) is considerably less than would be
expected given our inclusion of type I migration forces, however at
early times the protoplanetary masses are too small to be greatly
effected by inward type I drift and by the time they have grown an
order of magnitude to $\sim \mathrm{M}_\oplus$, inward drift remains
slow as the gas density in the inner system has fallen by an even
larger factor. The state of the solid components of the disk at the
starting point of Scenario~I shows little difference to its
equivalent case with no type I migration in \citet{fogg3}; however
by the time the system is aged to the starting point of Scenarios~IV
\& V, substantial inward movement of the protoplanetary population
has happened, depleting the outer disk and crowding the inner
regions with protoplanets (see Fig.~\ref{figure:2}). The adequacy of
an 8 day time-step for our late maturation runs is questionable as
significant material has moved inside 0.4~AU at $t > 1$~Myr.
However, protoplanetary growth in the inner part of our model disk
occurs rapidly and mostly at distances of $>$ 0.4~AU and tests we
conducted to examine this issue showed that disk forces prevent
strongly divergent behavior as a function of step size. The mass
distribution of our matured disks interior to 0.4~AU is therefore
not significantly affected by our choice of step size and, once the
giant planet is inserted at the start of each scenario, orbital
evolution is resolved with a shorter time-step of one tenth the
orbital period of the innermost object. As well as type I migration
generating edge effects at the inner edge of our solids disk, it is
likely that it would drive in other objects into our simulation
region from beyond 4~AU over this time: we have not included this
possibility here but comment on it further in
Sect.~\ref{discussion}.

\begin{table}
\caption{Disk data: after 0.1, 1.0 and 1.5 Myr of
evolution} %
\label{table:2}  %
\centering  %
\begin{tabular}{c| c c c}
 \hline\hline
Time~(Myr) & 0.1 & 1.0 & 1.5\\
Scenario ID & I & IV & V\\
 \hline
$M_{\mathrm{gas}}~(M_\mathrm{J})$ & 30 & 7 & 3\\
$M_{\mathrm{solid}}~(\mathrm{M}_{\oplus})$ & 24.8 & 24.8 & 23.5\\
$m_{\mathrm{max}}~(\mathrm{M}_{\oplus})$ & 0.55 & 1.07 & 1.83\\
$N$ & 51 & 24 & 16\\
$N'$ & 4087 & 2171 & 1585\\
$f_{\mathrm{proto}}$ & 0.20 & 0.47 & 0.51\\ %
 \hline\hline
\end{tabular}
\end{table}

The three aged disks summarized in Table~\ref{table:2} are used as
the basis for the type II giant planet migration scenarios studied
here. In each case, a giant planet of 0.5~$\mathrm{M_J}$ is inserted
at 5~AU after removing 0.4~$\mathrm{M_J}$ of gas from the disk
between 3--7~AU and the inner boundary of the simulation is reset to
0.014~AU, which is approximately the radius of a T-Tauri star
\citep{bertout}. The giant planet then proceeds to clear an annular
gap in the gas and migrates inward in step with its viscous
evolution. The simulations are halted when the giant reaches
hot-Jupiter territory at 0.1~AU: in Scenario I, this takes
$\sim$~114\,000 years, in close agreement with the viscous disk
evolution time at 5~AU, but in Scenarios IV and V the process takes
longer ($\sim$~152\,500 and 220\,000 years respectively) because at
their late times of starting most of the gas has been lost (see
Table~\ref{table:2}) and the remainder is less effective at driving
migration. The symplectic time-step for these runs was set to one
tenth the orbital period of the innermost object which was achieved
by dividing each simulation into a set of sequential sub-runs with
the time-step adjusted appropriately at each re-start. At late times
in each scenario when material had drifted interior to 0.1~AU, this
entailed working with time-steps of less than one day with overall
run times of several months as a consequence.

\begin{figure*}
\sidecaption
 \includegraphics[width=12cm]{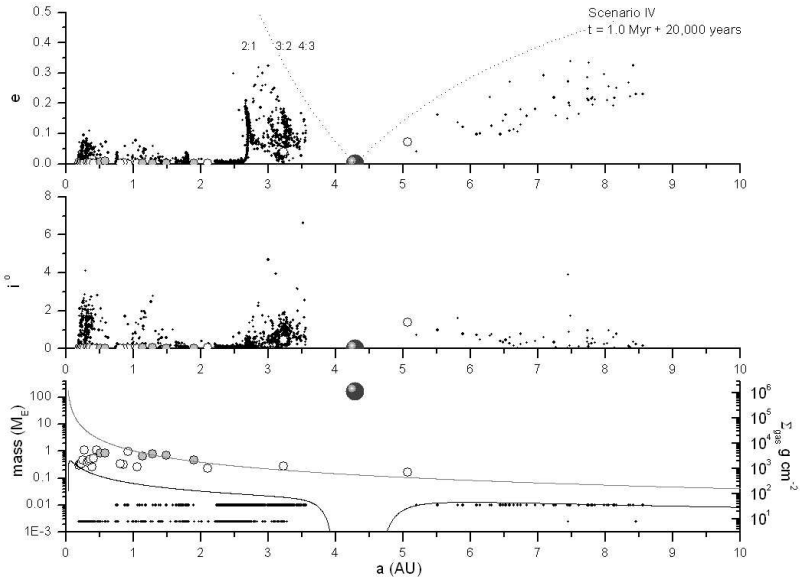}
 \caption{Scenario IV at $20\,000$ years after the start of giant planet
 migration, showing the mass, inclination and eccentricity of objects.
 Small black dots represent super-planetesimals; white
 filled circles are rocky protoplanets; grey filled circles are
 icy protoplanets and the large highlighted grey filled circle is
 the giant. Objects plotted between the dotted lines in the upper panel
 have orbits that intersect the orbit of the giant. The location of the
 2:1, 3:2 and 4:3 resonances with the giant are indicated. Gas surface
 density is read on the right hand axis of the lower panel, the upper
 grey curve being the unevolved profile at $t$ = 0 and the lower black
 curve being the current profile.}
 \label{figure:3}
\end{figure*}

\begin{figure*}
\sidecaption
 \includegraphics[width=12cm]{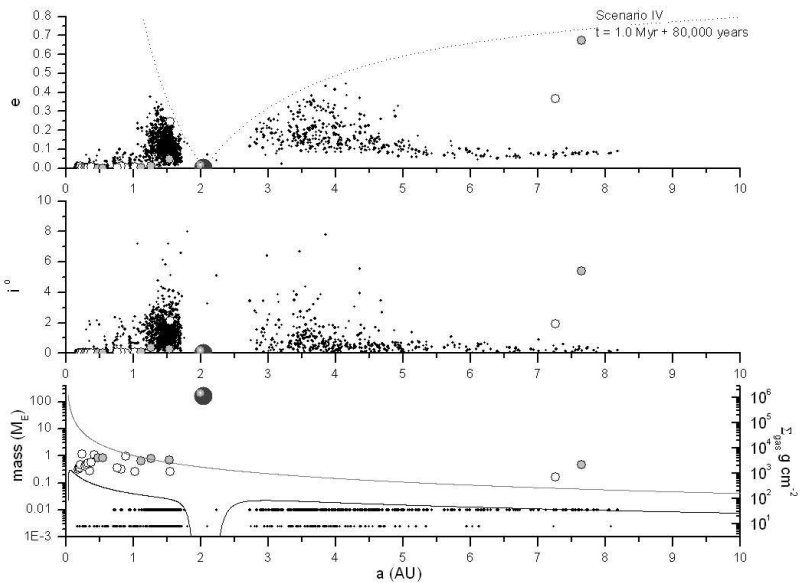}
 \caption{Scenario IV at $80\,000$ years after the start of giant planet
 migration. The giant has now moved inward to 2.04 AU and is beginning
 to enter the inner zone that has become packed with protoplanets due
 to previous type I migration.}
 \label{figure:4}
\end{figure*}

\begin{figure*}
\sidecaption
 \includegraphics[width=12cm]{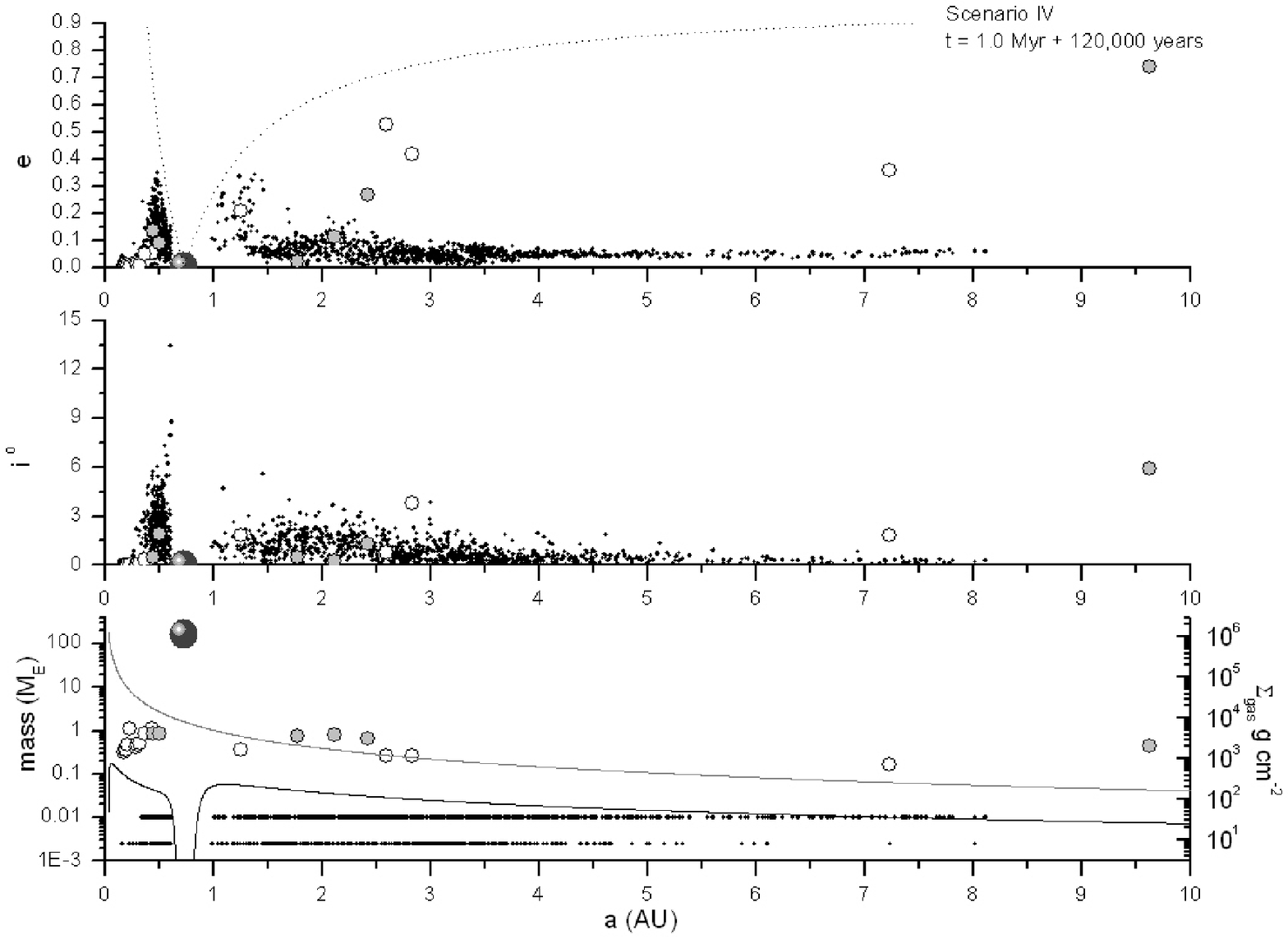}
 \caption{Scenario IV at $120\,000$ years after the start of giant planet
 migration. The giant has now moved inward to 0.72 AU. Six more protoplanets
 and a substantial number of planetesimals have been scattered into the
 external disk.}
 \label{figure:5}
\end{figure*}

\section{Results.}\label{results}

We describe first the behavior of one of our runs in detail and then
proceed to discuss the effect of adding type I migration to our
model by contrasting the results of our three scenarios presented
here with their counterparts in \citet{fogg3}.

\subsection{An account of Scenario IV}
\label{scenario4}

Scenario~IV is one of our "late" scenarios, taking place 1~Myr after
our model start time at a system age of $\sim$ 1.5~Myr. By this
point, and after removing the gas required for the giant planet's
envelope, the amount of gas remaining in the disk has fallen to just
$16\%$ of its initial value with the result that the giant takes
152\,000 years to migrate inward to 0.1~AU, longer than would be
predicted from the viscous evolution time. Four snapshots of the
evolution of Scenario IV are illustrated in Figs.~3 -- 6 showing the
mass, inclination and eccentricity of objects, and the gas surface
density vs. semi-major axis. The original provenance of the
protoplanets (interior or exterior to the snowline) is denoted by
the shading of its symbol as described in the caption to
Fig.~\ref{figure:3}. In the case of a merger between rocky and icy
protoplanets, this shading is determined by that of the most massive
of the pair.

\begin{figure*}
\sidecaption
 \includegraphics[width=12cm]{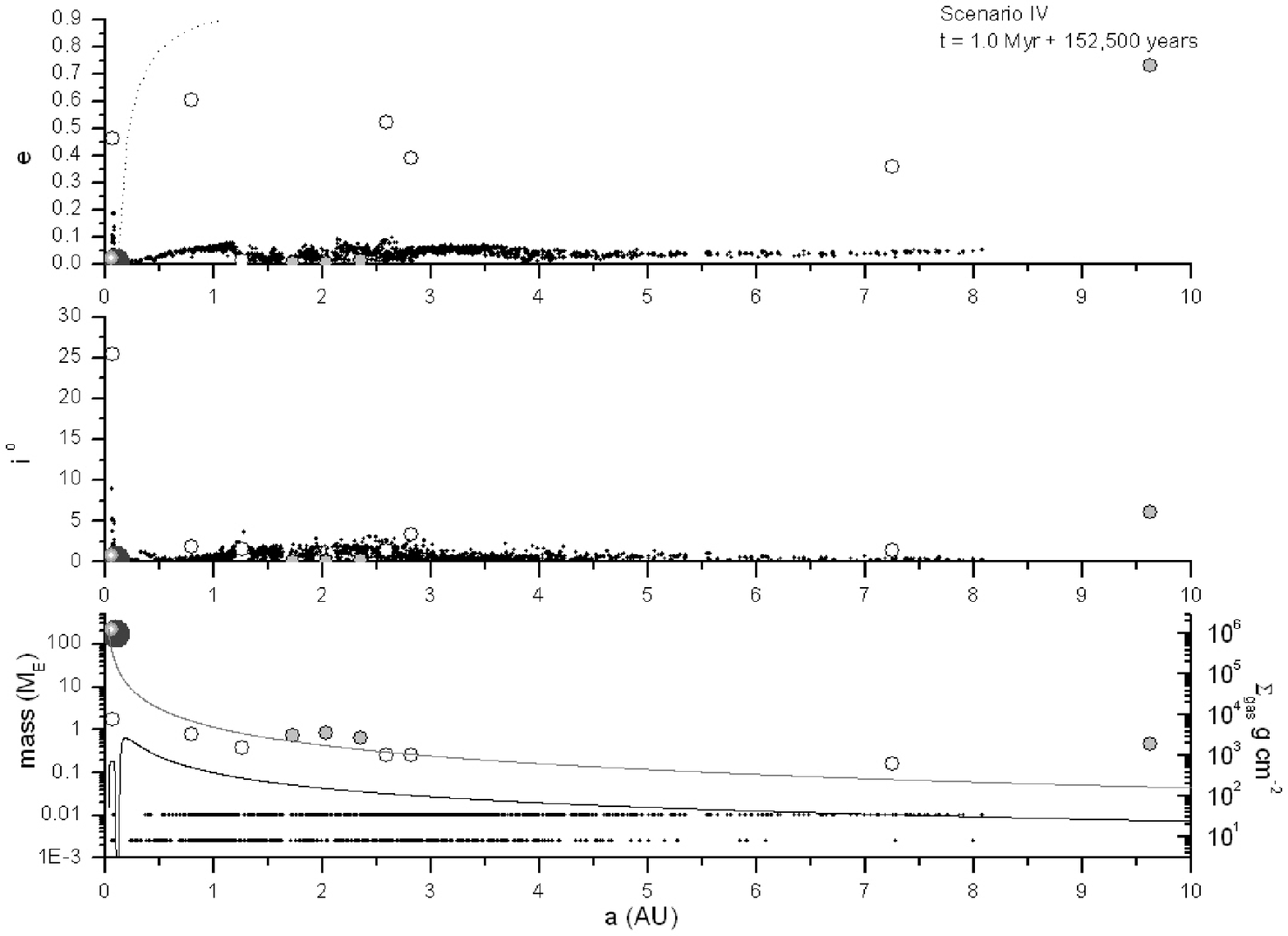}
 \caption{The end point of Scenario IV at $152\,500$ years after the
 start of giant planet migration. The giant is at 0.1 AU. About half
 of the original solids disk mass survives in external orbits and seven
 protoplanets are found between 0.7 -- 3~AU.
 A hot-Earth remains interior to the giant planet in an eccentric and
 inclined orbit.}
 \label{figure:6}
\end{figure*}

An early stage in the run at 20\,000 years after the introduction of
the giant planet is shown in Fig.~\ref{figure:3}. The giant has
moved in to 4.29~AU and has caused significant excitation of the
outer solids disk material at first order resonances. Noticeable
also is that the outer disk beyond $\sim$ 2~AU has been largely
cleared of protoplanets due to previous type I migration. Only two
low mass protoplanets remain in this region, one of them having
recently become caught at the interior 3:2 resonance with the giant
planet and the other having already been scattered into an external
orbit.

An intermediate stage in Scenario~IV at 80\,000 years is shown in
Fig.~\ref{figure:4} where the giant planet has migrated inward to
2.04~AU and is at the point of entering the zone that is crowded
with protoplanets at $a < 2$~AU. Most of the planetesimals
originating in the outer disk have now been shepherded interior to
2~AU, greatly increasing the surface density of solids caught
between the 2:1 and 4:3 resonances. Two protoplanets have been
captured at the 3:2 resonance and are in the process of having their
orbits excited to an eccentricity where they will cross the giant
planet's orbit and become vulnerable to scattering or accretion.
Type I eccentricity damping is therefore not always strong enough to
counteract the resonant pumping exerted by the giant planet as is
shown by the additional $0.46~\mathrm{M}_\oplus$ protoplanet that
has been expelled into the growing scattered disk. One reason for
this is that $t_e$ increases with the cube of eccentricity
(Eq.~\ref{ecctime}) greatly reducing the resistance of type I
eccentricity damping to strong perturbations. It is also evident by
this stage that the giant planet is easily capable of catching up
and overtaking the protoplanets which have only moved inward
slightly in the meantime. For example, the type I migration time
given by Eq.~\ref{migtime} for a $0.5~\mathrm{M}_\oplus$ protoplanet
at 2~AU, with $e = 0$, $h/r = 0.056$ and $\Sigma_\mathrm{g} =
150~\mathrm{g~cm^{-2}}$ (a twelve-fold reduction in gas surface
density at this location from that at $t = 0$) is $t_\mathrm{m}
\approx 6$~Myr, comfortably larger than the type II migration time
of the giant.

Fig.~\ref{figure:5} shows the system 120\,000 years after the
introduction of the giant planet which has now reached 0.72~AU. An
external disk is building up. Eight more protoplanets have had close
encounters with the giant planet, resulting in two of them being
accreted and six undergoing external scattering. The orbits of the
most massive and least eccentric of these bodies are seen to be
damping fast, as would be expected from the influence of
Eq.~\ref{ecctime}. Some of this damping might also be due to
dynamical friction as a substantial number of super-planetesimals
have also been scattered into the same vicinity. Most of the solids
mass however remains in the shepherded disk fraction, interior to
the giant planet. Gas drag and type I damping forces in this region
however are now two orders of magnitude weaker than a nominal
$3\times\mathrm{MMSN}$ model as gas accretion onto the central star
has reduced $\Sigma_\mathrm{g}$ here by a factor of $\sim$ 25 since
$t = 0$ and dissipation of spiral waves generated by the approaching
giant have recently reduced $\Sigma_\mathrm{g}$ by an additional
factor of $\sim$ 4.

\begin{table*}
\caption{Fate of the disk mass at the end of each run: results for
scenarios without ($\nrightarrow$) and with ($\rightarrow$) type I migration.} %
\label{table:3}  %
\centering  %
\begin{tabular}{c| c c| c c| c c}
 \hline\hline %
Scenario & I$\nrightarrow$ & I$\rightarrow$ & IV$\nrightarrow$ & IV$\rightarrow$ & V$\nrightarrow$ & V$\rightarrow$\\
 \hline\hline %
Total Initial Solids $(\mathrm{M}_{\oplus})$ & 24.81 & 24.81 & 24.81
& 24.81 & 24.81 & 23.47\\
 \hline %
Total Surviving Solids $(\mathrm{M}_{\oplus})$ & 16.60~(67\%) &
18.70~(75\%) & 21.23~(86\%) & 15.13~(61\%) & 20.22~(81\%) &
18.23~(78\%)\\
 \hline %
Interior Surviving Solids $(\mathrm{M}_{\oplus})$ & 0.88~(4\%) &
3.44~(14\%) & 0.84~(3\%) & 1.84~(7\%) & 0.31~(1\%) & 6.23~(27\%)\\
$N, f_{\mathrm{proto}}$ & 0, 0 & 1, 0.99 & 0, 0 & 1, 0.96 & 0, 0 & 1, 1\\
 \hline %
Exterior Surviving Solids $(\mathrm{M}_{\oplus})$ & 15.72~(63\%) &
15.27~(62\%) & 20.39~(82\%) & 13.29~(54\%) & 19.90~(80\%) & 12.00~(51\%)\\
$N, f_{\mathrm{proto}}$ & 39, 0.27 & 34, 0.22 & 31, 0.63 & 9, 0.34 &
23, 0.66 & 7, 0.24\\
 \hline %
Accreted by Star $(\mathrm{M}_{\oplus})$ & 0.01~(0.04\%) & 0.0~(0\%)
& 0.0~(0\%) & 0.0~(0\%) & 0.0~(0\%) & 0.01~(0.04\%)\\
 \hline %
Accreted by Giant $(\mathrm{M}_{\oplus})$ & 8.20~(33\%) &
6.01~(24\%) & 3.41~(14\%)
& 9.68~(39\%) & 4.59~(19\%) & 5.23~(22\%)\\
 \hline %
Ejected $(\mathrm{M}_{\oplus})$ & 0.00~(0\%) & 0.1~(0.4\%) &
0.17~(1\%) &
0.0~(0\%) & 0.0~(0\%) & 0.0~(0\%)\\
 \hline\hline %
\end{tabular}
\end{table*}

\begin{figure}
 \resizebox{\hsize}{!}{\includegraphics{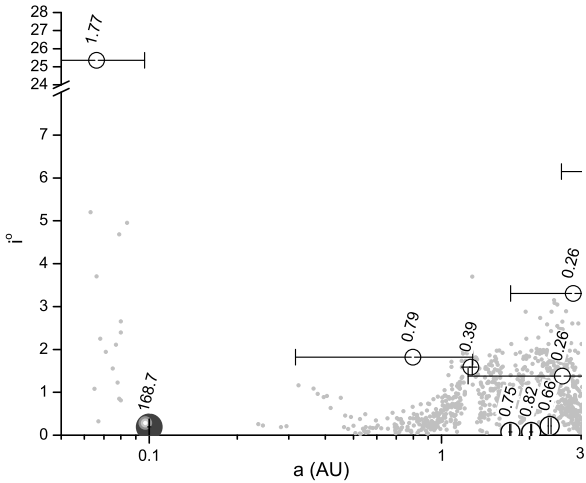}}
 \caption{Scenario IV interior to 3 AU at the point where the giant planet
 reaches 0.1 AU. Inclination is plotted vs semi-major axis. Super planetesimals
 are the grey dots and protoplanets the open circles. Planets also have marked
 their masses in $\mathrm{M}_\oplus$ and their radial excursions due to orbital
 eccentricity.}
 \label{figure:7}
\end{figure}

Scenario~IV ends at 152\,500 years after the introduction of the
giant planet when it reaches 0.1~AU and the system configuration at
this point is shown in Fig.~\ref{figure:6}. The final 0.6~AU of the
giant's migration has had a destructive effect on the densely packed
inner system material. Of the twelve protoplanets remaining interior
to the giant at the point shown in Fig.~\ref{figure:5}, there are
only two remaining: a $0.79~\mathrm{M}_\oplus$ planet scattered to
0.8~AU, and a $1.77~\mathrm{M}_\oplus$ hot-Earth candidate at
0.062~AU, close to the 2:1 resonance, with an eccentric, inclined
orbit ($e = 0.46, i = 25.3^\circ$). The other ten, which were poorly
damped due to low gas densities, and compacted into an ever
shrinking volume, underwent a series of violent mutual impacts
followed by eventual accretion by the giant planet. A close
encounter with the last of these hot-Earths to hit the giant was
responsible for the high inclination of the orbit of the one
survivor which, at $i = 25.3^\circ$, is too low for the Kozai
mechanism to play a role in subsequent orbital evolution. Despite
these losses however, about half the mass of the original solids
disk survives in the scattered external disk and most of this at
radial distances $0.7 < r < 3$~AU (see the magnified illustration in
Fig.~\ref{figure:7}). Seven substantial protoplanets ranging in mass
from $0.26 - 0.82~\mathrm{M}_\oplus$ now occupy this region along
with a large quantity of yet to be accreted planetesimals. Type I
eccentricity damping has already circularized the orbits of four of
the protoplanets which might be expected to undergo some degree of
renewed inward type I migration before complete gas dispersal
followed by a final phase of planetesimal clear-up and giant
impacts. Allowing for this, the configuration shown in
Fig.~\ref{figure:7} suggests a good chance that the final planetary
arrangement here could include a planet in the system's habitable
zone and we return to this issue in Sect.~\ref{outer-planets}.

\subsection{Comparison with simulations not including type I
migration.}\label{comparison}

Data showing the fate of the solids disk mass at the end of the
three scenarios presented here that include type I migration are
given in Table~\ref{table:3} in the columns headed with a
$\rightarrow$ symbol and are compared against their counterparts
from \citet{fogg3} in the columns headed with a $\nrightarrow$
symbol. Each column gives the total mass of solid material in the
disk at the start of giant planet migration; the mass and the
percentage of the initial disk surviving at the end of the
migration; and the mass surviving interior and exterior to the giant
planet's final orbit at 0.1~AU, including the number of remaining
protoplanets and the protoplanetary mass fraction in each respective
partition. The bottom three rows give the data for the mass loss
channels which are accretion by the central star, accretion by the
giant planet and ejection into a hyperbolic orbit.

These data show that a large fraction of the disk mass survives the
passage of the giant planet in each case: inclusion of type I
migration at the nominal rate does not result in the inner system
being cleared of planet-forming material. In the two versions of
Scenario~I slightly more mass remains with type I migration
operating than without. However, this is largely due to the presence of a
$3.42~\mathrm{M}_\oplus$ hot-Earth inside 0.1~AU captured in the 3:2
resonance with the giant planet with an eccentricity of 0.29.
Generally the data for Scenarios~I$\nrightarrow$ and I$\rightarrow$ are
similar, suggesting that type I migration only has a small influence
on the outcome of giant planet migration through a young inner
system disk where protoplanetary masses remain small. In our more
mature Scenarios~IV \& V we do find less solids mass surviving when
type I migration is included, but this loss is not catastrophic:
61\% of the solids disk remains at the end of Scenario~IV (see
Figs.~6 \& 7) and 78\% remains at the end of Scenario~V. Almost
exclusively, the mass that is lost is accreted by the giant planet.
The increase in efficiency of this loss channel results from the
fact that, due to previous inward type I migration, protoplanetary
encounters with the giant planet tend to occur at smaller orbital
radii where the probability of accretion versus scattering is
relatively enhanced. In Scenario~V this additional mass loss is
minor as much of the inner system material had by this late time
accreted into substantial protoplanets which were able to migrate
ahead of the giant inside 0.1~AU: a situation that can be seen
developing in Fig.~\ref{figure:2}. These eventually co-accreted to
form a $6.23~\mathrm{M}_\oplus$ hot-Earth that was captured into the
2:1 resonance with the giant planet near the end of the simulation
exciting its orbit to an eccentricity of 0.41. Future accretion of
this body by the giant planet is a possibility and we examine the
issue of the long-term survival of our hot-Earth planets in
Sect.~\ref{hotearths}.

Across the board in Table~\ref{table:3}, the consistent trends that
emerge when type I migration is added to our model are:

\begin{enumerate}
\item \emph{Shepherding is modestly enhanced.} This can be seen from
the larger percentage of the original solids disk mass that is found
in the interior surviving solids fraction, which includes in each
case a single hot-Earth type planet with an eccentric orbit captured
at a first order resonance with the giant planet. Such objects were
seen to form, but not to survive, in our analogous runs in
\citet{fogg3}: their typical fate being accretion by the giant
planet. The hot-Earths generated here have narrowly escaped
this fate themselves and their long-term survival would depend on
the cessation of any further inward migration of the giant and the
effectiveness of any continuing type I eccentricity damping exerted
by the tenuous gas (reduced by over two orders of magnitude from a
$\Sigma_g \propto r^{-1.5}$ model) remaining in the planet's
vicinity. We do not recover the well-damped hot-Neptune or hot-Earth
type planets generated in \citet{fogg1} where we assumed a steady
state gas disk with a fixed $\Sigma_g \propto r^{-1.5}$ surface
density profile. In this case, the persistently high gas densities
caused more rapid orbital decay of planetesimals, strong dynamical
friction and high rates of collisional damping which were sufficient
to drag several protoplanets into low orbits where a rapid final
phase of accretion was often sufficient to free the remaining body
from resonant locking with the giant planet. Here, our evolving and
decaying gas disk weakens both dynamical friction and type I
eccentricity damping; whilst type I orbital decay facilitates inward
movement of protoplanets, at late times it also brings about a
separation of protoplanets from the bulk of planetesimals, further
reducing the influence of dynamical friction on the innermost
bodies. The overall dissipation in our present model therefore only
results in a maximum of 27\% of the original solids disk partitioned
into the inner remnant.

\begin{figure}
 \resizebox{\hsize}{!}{\includegraphics{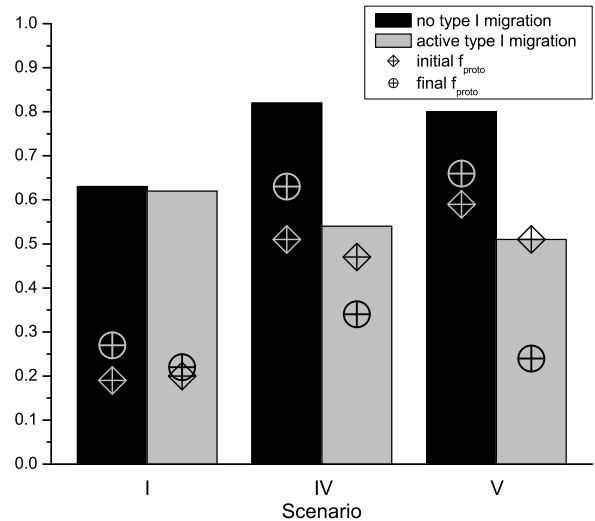}}
 \caption{Mass fractions of scattered disks generated in the present
 work compared to those of analogous scenarios not including type I
 migration and before and after protoplanet mass fractions.}
 \label{figure:8}
\end{figure}

\item \emph{Scattering is modestly reduced and strongly selective.}
Our simulations that include type I migration produce less massive
scattered disks, but not dramatically so. A stronger effect is that
the likelihood of external scattering is now biassed in favor of
planetesimals, generating immature exterior solids disks with fewer
protoplanets. This can be seen in Fig.~\ref{figure:8} where the
black bars show the mass of the scattered disk resulting from runs
with no type I migration and the grey bars show those resulting from
the present work. The superimposed lozenge symbols show the disk
protoplanet mass fraction ($f_{proto}$) at the start of each
scenario and the circle symbols give the value of $f_{proto}$ at the
end. In late scenarios the addition of type I migration to our model
has resulted in the reduction of the mass of the outer disk remnant
by $\sim 30\%$ which still leaves $\sim 50\%$ of the mass of the
original ending up in the scattered disk. These disks however are
less mature in the sense that they contain a greater fraction of
small bodies. In \citet{fogg3} it was noted that one effect of giant
planet migration through a solids disk was to increase $f_{proto}$
in the external remnant by selectively scattering protoplanets.
Fig.~\ref{figure:8} shows that in late scenarios, this trend is
reversed here as type I migration, especially at late times, causes
protoplanets to be preferentially shepherded: encounters with the
giant tend to occur closer in so that fewer of them escape into the
external remnant allowing the scattered planetesimal population to
dominate. The external disks generated here however, though of
reduced mass, should still be capable of supporting future planet
formation. The protoplanets they contain can be of substantial mass
and are typically emplaced at smaller semi-major axes and the more
abundant planetesimal population and residual type I migration
forces provide for stronger damping of protoplanets after scattering
has occurred.
\end{enumerate}

\subsection{Post--migration terrestrial planet formation}\label{outer-planets}

\begin{figure*}
\sidecaption
 \includegraphics[width=12cm]{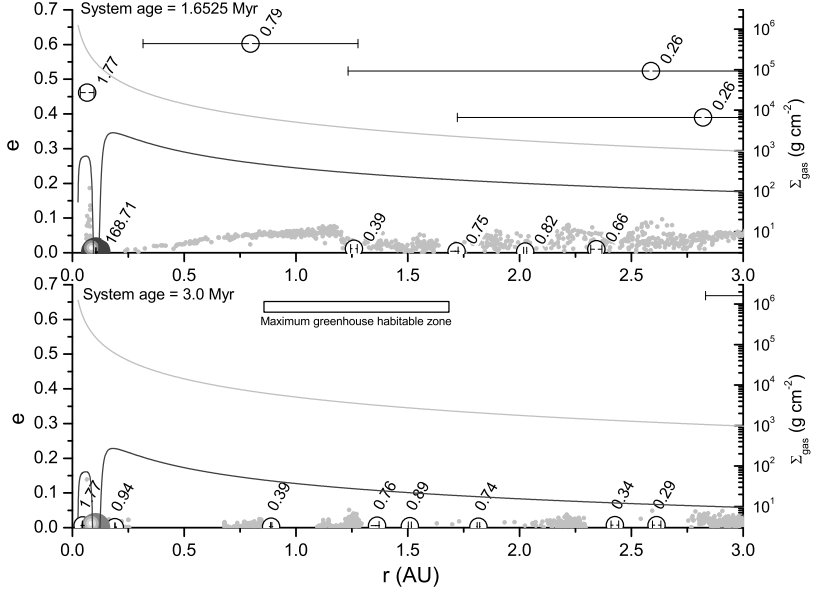}
 \caption{Details of the inner 3~AU at the end of Scenario IV (top panel)
 and 1.3475~Myr later at the point of gas dispersal. Eccentricity is
 plotted against semi-major axis with orbital excursions due to
 eccentricity illustrated with error bars. Planetary masses are labeled
 in units of $\mathrm{M}_\oplus$. $\Sigma_\mathrm{g}$ is read from the right
 hand axes: the upper curve being the profile at $t = 0$
 (system age = 0.5 Myr) and the lower curve being the present profile.
 The extent of the maximum greenhouse habitable zone (0.8 - 1.7~AU)
 is shown by the horizontal bar.}
 \label{figure:9}
\end{figure*}

To model the completion of planet formation in our scattered disks
would require simulating a further $\sim$ 100~Myr of accretion and
is beyond the scope of this paper. Shorter range integrations
however are still of considerable interest as type I migration
forces only decline to zero when all the remaining local gas has
gone. In this section, we continue the evolution of Scenario IV to a
system age of 3~Myr, requiring an additional 1.3475~Myr of simulated
run time, at which point we assume all residual gas is rapidly lost
to photoevaporation. Over this period the mass of the gas disk falls
from $\sim 5.5~\mathrm{M_J}$ to $\sim 0.55~\mathrm{M_J}$. In order
to evade the severe limitations of our viscous disk algorithm
time-step, which adapts to a tiny value with the giant planet at
$\sim$ 0.1~AU, we turn off the type II migration forces effecting
the hot-Jupiter and replace our viscous disk with an exponentially
declining disk that maintains the previous surface density profile
and declines with a mass e-folding time of 581\,514 years. We evolve
the system henceforth with a 2 day N-body time-step. This set-up
assumes an unspecified migration halting mechanism on the part of
the hot-Jupiter that is not fully consistent with our model and we
comment on the consequences of this in Sect.~\ref{discussion}.

\begin{figure}
 \resizebox{\hsize}{!}{\includegraphics{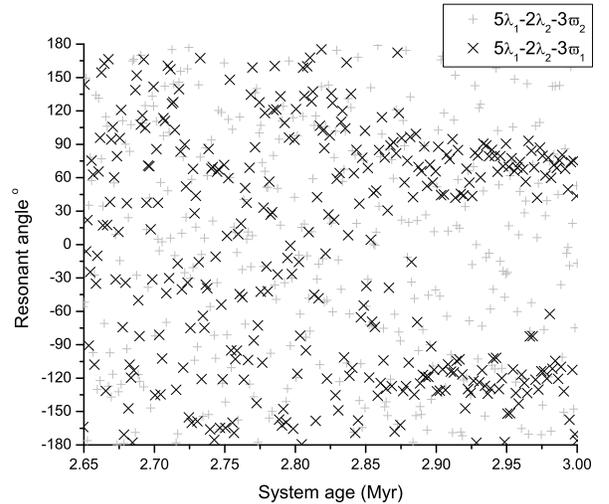}}
 \caption{Resonant angles for the 5:2 mean motion resonance between
 the giant planet and the innermost external protoplanet. Resonant
 capture is indicated $\sim$ 125\,000 years before the stage shown in
 the bottom panel of Fig.~\ref{figure:9}.}
 \label{figure:10}
\end{figure}

\begin{figure}
 \resizebox{\hsize}{!}{\includegraphics{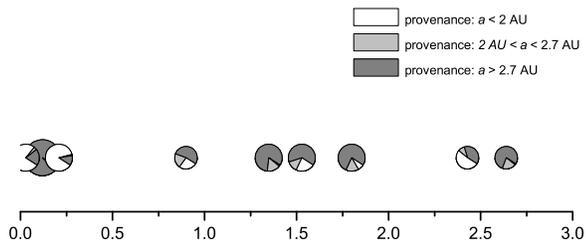}}
 \caption{Material composition of protoplanets in the extended
 Scenario IV at the stage shown in
 the bottom panel of Fig.~\ref{figure:9}.}
 \label{figure:11}
\end{figure}

\begin{figure*}
\sidecaption
 \includegraphics[width=12cm]{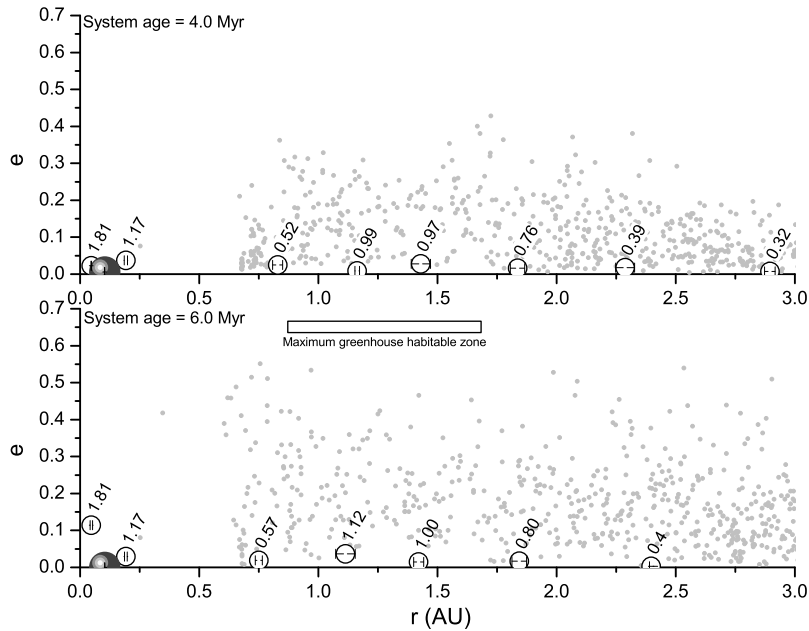}
 \caption{Details of the inner 3~AU of the Scenario~IV system after
 3~Myr of gas-free dynamics. The top and bottom panels show
 configurations at system ages of 4 and 6~Myr respectively.}
 \label{figure:12}
\end{figure*}

Details of the evolution of the inner 3~AU of the system during this
terminal phase of gas depletion are shown in Fig.~\ref{figure:9}
where the top panel shows the state of play at the end of Scenario
IV and the bottom panel shows it at 1.3475~Myr later at a system age
of 3~Myr. During this period we find that dynamical friction and
residual type I migration forces rapidly damp and circularize the
orbits of protoplanets with high eccentricities. Gas drag is also
sufficient to maintain the eccentricities of planetesimals at a low
value. The four well damped protoplanets between 1.25 -- 2.5~AU in
the upper panel of Fig.~\ref{figure:9} grow by a few percent from
the surrounding planetesimal field and migrate inward an additional
$\sim$ 0.6~AU. By the time we remove the remaining gas in the bottom
panel, two more planets have moved into the maximum greenhouse
habitable zone \citep{kasting} giving a total of three planets sited
between 0.8 -- 1.7~AU. The migration of these four protoplanets
shows some of the convoying behavior described in \citet{mcneil}:
convergent migration can cause protoplanets to bunch together and
lock into a stable arrangement of mean motion resonances whereupon
the pattern then migrates inward at some average drift rate. In the
bottom panel of Fig.~\ref{figure:9} the 0.76~$\mathrm{M}_\oplus$
planet and its 0.89~$\mathrm{M}_\oplus$ neighbour are caught at
their 7:6 resonance and the outermost of this pair is close to the
4:3 resonance with its outer 0.74~$\mathrm{M}_\oplus$ neighbour. The
innermost external protoplanet, having only recently been scattered,
starts with $m_{\mathrm{proto}} \simeq 0.78~\mathrm{M}_\oplus$, $a
\simeq 0.8$~AU and $e \simeq 0.6$ but damps rapidly and migrates
inward, growing to 0.94~$\mathrm{M}_\oplus$ and eventually halting
at $a = 0.19$~AU at the 2:5 resonance with the hot-Jupiter (see
Fig.~\ref{figure:10}). This behavior is an example of the tendency,
when convergent migration is in play, for giant planets to collect
smaller bodies at external mean motion resonances \citep{thommes4}.

Our model also tracks the volatile composition of protoplanets,
recognizing three crude material provenances: material originating
at $a < 2$~AU is assumed to be dry rock; from between $2~\mathrm{AU}
< a < 2.7~\mathrm{AU}$ its water content is assumed to be similar to
that of chondritic asteroids; and from beyond $a > 2.7$~AU it is
assumed to be water-rich trans-snowline material. Previous models
have demonstrated how a migrating giant planet drives volatile-rich
material into its inner system, boosting the potential water content
of any planets that form there \citep{raymond2,fogg3,mandell}. This
trend is accentuated when type I migration forces are included as
volatile-rich protoplanets can now migrate directly inward. In
Fig.~\ref{figure:11} we show the composition of the protoplanets in
the extended Scenario~IV at a system age of 3~Myr. All the bodies
have a substantial endowment of outer solids disk material
especially those at $a > 0.8$~AU. The largest three of these
planets, the resonant trio situated between 1.3 -- 1.9~AU, all
originated from beyond the snowline at $a > 2.7$~AU and accreted
copious volatile-rich material during their inward journey. Ocean
worlds are a probable outcome of this kind of evolution
\citep{kuchner,leger}.

We have evolved this system, in the absence of gas, for a further
3~Myr to a system age of 6~Myr and two configurations over this
period are shown in Fig.~\ref{figure:12}. The cessation of type I
migration forces causes a shift to a more oligarchic style of growth
\citep{kokubo1}. Orbital repulsion breaks up the resonant convoy
showing in Fig.~\ref{figure:9} and the entire protoplanet chain
dynamically expands. The cessation of gas drag results in
planetesimal orbits becoming much more excited in the process of
exerting dynamical friction. Although this reduces protoplanetary
accretion cross-sections, protoplanets no longer shepherd gaps in
the solids disk and continuing growth is in progress.

\begin{figure*}
\sidecaption
 \includegraphics[width=12cm]{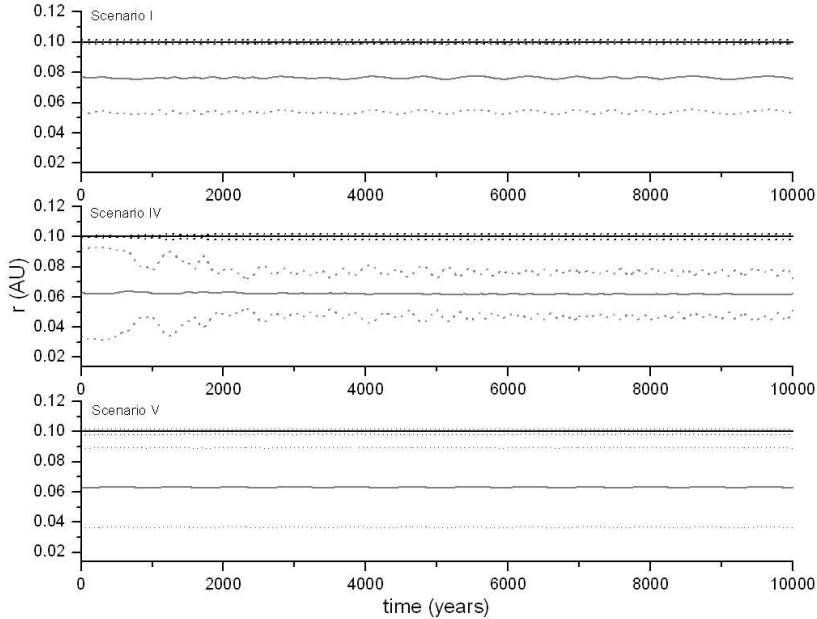}
 \caption{Further evolution of the orbits of the interior planets resulting from
 the three scenarios. Solid lines are semi-major axes, dotted lines are
 periastra and apastra. Black represents the giant planet and grey
 represents the hot-Earth.}
 \label{figure:13}
\end{figure*}

The final state of the system shown in the bottom panel of
Fig.~\ref{figure:12} cannot be predicted in detail. We speculate
that planetesimals will eventually be accreted or cleared, with some
small fraction of them possibly surviving in the prominent gap in
the planetary distribution at $\sim$~0.5~AU. Future mergers of
protoplanets via giant impacts are less certain as they all lie in
near-circular and well-spaced orbits: the closest being the two
bodies between 1 -- 1.5~AU which are nonetheless separated by
$\sim$~19 mutual Hill radii. We expect therefore that the final
configuration of the Scenario~IV system will contain either one or
two Earth-mass planets in its habitable zone and longer range
integrations of this system are underway.

\subsection{Hot-Earth survival}\label{hotearths}

Having used a 2 day time-step, the additional simulation represented
by Fig.~\ref{figure:12} does not adequately resolve the orbital
evolution of the hot-Earth planet at 0.062~AU. Single hot-Earths,
captured at interior first order mean motion resonances resulted
from all our scenarios presented here, but their eccentric orbits
with apastra close to the radial distance of the giant planet raise
questions over long term survival. We have examined this issue by
running the orbits of our hot-Earth candidates for a further 10\,000
years ($\sim 10^6$ orbits at 0.05~AU), with a half day time-step, in
the absence of gas and including the giant planet and any remaining
debris interior to 0.1~AU. This assumes the existence of an interior
gaseous cavity which is not fully consistent with our scenario end
points and we comment on this further in Sect.~\ref{discussion}.

\begin{figure}
 \resizebox{\hsize}{!}{\includegraphics{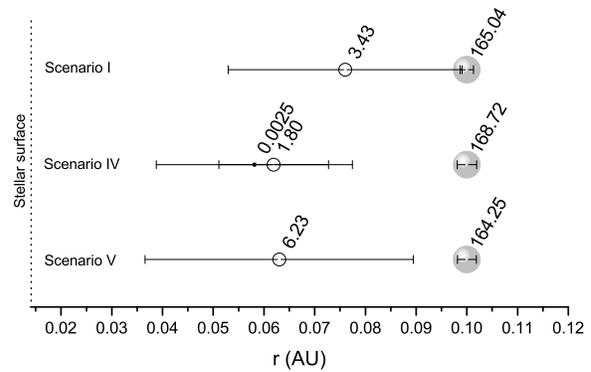}}
 \caption{Hot-Earth -- hot-Jupiter systems resulting from an
 additional 10\,000 years of simulation. Error bars show the
 radial distance between periastron and apastron. Numeric
 labels give masses in $\mathrm{M}_\oplus$.}
 \label{figure:14}
\end{figure}

The orbital evolution resulting from these extended runs is
illustrated in Fig.~\ref{figure:13} and their final configuration is
shown in Fig.~\ref{figure:14}. All the interior planets remain. The
least convincing as a longer-term survivor is the
3.43~$\mathrm{M}_\oplus$ hot-Earth present in Scenario~I: it
accreted very little additional debris (which was accreted by the
giant instead) and continues to orbit at the 3:2 resonance with an
eccentricity of $\simeq$ 0.3, bringing it very close to crossing the
giant's orbit at apastron. Close encounters have so far been avoided
by the orbital phasing resulting from the resonance, but we have not
investigated how long this state could persist. In the case of
Scenario~IV, we start with a 1.77~$\mathrm{M}_\oplus$ hot-Earth at
the 2:1 resonance in an excited orbit with $e = 0.46$ and $i =
25^\circ$, but this planet succeeds in accreting an extra
0.03~$\mathrm{M}_\oplus$ of debris, cleaning up all of it but the
one remaining super-planetesimal showing in Fig.~\ref{figure:14},
with the result that its eccentricity is damped to $e \simeq 0.2$ by
the end of the run period. Its inclination is not damped and one
accretion event increases it slightly to $i \simeq 30^\circ$ but the
hot-Earth and the hot-Jupiter are now well separated and long-term
survival looks promising. In Scenario~V there was no debris for the
hot Earth to clean up and Fig.~\ref{figure:13} shows that its orbit
remains unchanged over 10\,000 years. This 6.23~$\mathrm{M}_\oplus$
planet remains close to the 2:1 resonance in a very low inclination
orbit with $e \simeq 0.41$ which at no point comes dangerously close
to the orbit of the giant planet. This hot-Earth is also a possible
long-term survivor, especially as stellar tidal forces would be
expected to gradually circularize its orbit over much longer
timescales.

\section{Discussion.}\label{discussion}

Any simulation of planet formation is restricted by its assumptions
and incomplete input physics. Below, we discuss three areas where future
improvements are possible as well as the caveats they might hold for
our present results. \\
({\it i}) {\em Migration halting mechanisms.} \hspace{1mm} In all
our simulations we have halted the migration of our giant planet at
0.1~AU where we assume that whatever mechanism is responsible for
stranding hot-Jupiters in their final orbits comes into play. This
mechanism is unknown and proposed explanations include the presence
of a central magnetospheric cavity in the gas which could decouple
an intruding giant planet from the disk \citep{lin2}; another is
that of fortuitous gas disk dispersal where the eventual loss of the
gas disk strands those migrating giant planets that still remain at
arbitrary orbital radii \citep{trilling}. The former
could apply to any of our scenarios, early or late, whilst the
latter is more appropriate to late scenarios when most of the gas
has already dispersed. Neither of these mechanisms are fully
consistent with our present model. If we were to run it further, we
would expect the giant planet to come to rest somewhere between
0.02--0.1~AU but we have yet to model any of the physics of a
magnetospheric cavity and time-step limitations that stem from our
viscous disk algorithm make it impractical to simulate further type
II migration interior to 0.1~AU. The only difference that stopping
the hot-Jupiter closer to the star would make to our results is that
more compact systems would be less likely to host interior hot-Earth
or hot-Neptune type planets. Further resonant shepherding of such
objects inward of 0.1~AU runs the risk of driving them into the
central star \citep{fogg1}. Our results for scattered disk
characteristics and their prospects of renewed terrestrial planet
formation would be unaffected by the final fate of the hot-Jupiter.
If fortuitous disk dispersal is assumed to be the halting mechanism,
then this could be simulated in our model by including a
photoevaporation algorithm which removes gas at some small and
steady rate that accords with observations. This would result in a
rapid dispersal of the of the remaining gas at late times
\citep[e.g.][]{clarke,alexander} and a termination of type II
migration. Work is ongoing to include this mechanism in our model
and to conduct simulations where giant planets strand at a variety
of distances in a fully self-consistent way. \\
({\it ii}) {\em Supply of outer disk material.} \hspace{1mm} In our
previous models the radial drift of protoplanets is due to dynamical
stirring and a degree of coupling to the slow inward movement of
planetesimals via dynamical friction. In these cases it was
reasonable to assume that our giant planet formed at 5~AU and that
any supply of extra material over our run times into the inner disk
from beyond that distance would be insignificant. However the
inclusion of type I migration makes the heaviest and most evolved
objects more inwardly mobile. As shown in Fig.~\ref{figure:2}, our
late matured disks are largely cleared of protoplanets outward of
$\sim$2~AU, making one wonder if this region could have been
repopulated in the interim by objects which migrated from beyond
4~AU. This seems plausible as the giant planet's core must also have
migrated so that we can no longer propose consistently that it
accreted its core at 5~AU. We have not yet attempted to model the
extent of any such addition of external mass to our solids disk but
it is clear from our results that any additional protoplanets
encountered by the giant between 2--4~AU are more likely to be
scattered than accreted. Thus, any late supply of large bodies is
likely to contribute to the population of the scattered disk
boosting both its mass, maturity ($f_{proto}$), and volatile
content. \\
({\it iii}) {\em Planetesimal size evolution.} \hspace{1mm} We have
assumed a uniform population size of 10~km planetesimals which
results in an identical effect of gas drag on any planetesimal at a
particular orbital distance. In reality, but more challenging to
adapt to a $N + N'$-type simulation, we would expect an evolving
planetesimal size spectrum and a much more varied response of small
objects to gas drag. In areas of the solids disk where planetesimals
are strongly stirred, such as between the 2:1 and 4:3 resonances
with the giant planet, high energy mutual collisions might reduce
planetesimals down to smaller chunks of rubble. This rubble would be
subject to stronger dissipation from the gas resulting in it being
shepherded inward of the giant more efficiently. In \citet{fogg3} we
suggested that much of this material should still end up in the
scattered disk as it would be expected to accrete very efficiently
onto inner protoplanets (see \citet{chambers4}) that would be
scattered later. This argument is weaker when type I migration is
included. Supplying inner planets with an enhanced feedstock would
not only boost their growth rates, but also their inward drift. We
have shown in Sect.~\ref{comparison} how the influence of type I
migration results in fewer protoplanets surviving in external orbits
so boosting inward migration rates by enhancing mass growth would
accentuate this trend. However, as accretion rates depend as
$\dot{m}_{proto} \propto \Sigma\Omega$ we would expect that the bulk
of this additional growth would take place at late times when the
giant planet has compacted the shepherded fraction within a small
orbital radius. It seems unlikely therefore that the additional
dissipation resulting from planetesimal fragmentation would prevent
entirely the scattering of some large bodies into the external disk.
\\
({\it iv}) {\em Initial nebular surface density profile.}
\hspace{1mm} In common with many other studies of terrestrial planet
formation, our models adopt an initial condition of a MMSN-type
protoplanetary disk with an $r^{-1.5}$ surface density profile
\citep{weidenschilling,hayashi}. Other models however predict that a
shallower profile ($\Sigma \sim r^{-1}$) may be more realistic
\citep[e.g.][]{davis,andrews}. We have not yet explored the effect
of running our present model with a shallower initial profile but
expect that it would not drastically change our results. Previous
models that have explored this issue
\citep[e.g.][]{chambers5,raymond1b,mcneil} find that the
architectures of their resulting terrestrial planetary systems
depend only weakly on the slope of the initial surface density set
up with shallower gradients tending to result in more substantial
planets forming further from the central star. If a giant planet
were to migrate through such a disk, we expect that more mass would
be encountered earlier in the migration, resulting in a more massive
scattered disk and hence even more favorable prospects for external
terrestrial planet formation.

\section{Conclusions.}\label{conclusions}

Previous models of a giant planet migrating through an inner system
protoplanetary disk have all predicted that the process generates a
scattered external disk massive enough to support renewed planet
formation \citep{fogg1,fogg2,raymond2,fogg3,mandell}. None of these
works included the possible action of type I migration forces
exerted by the gas which would cause heavier bodies to drift
progressively closer to the central star. Since this process would
clearly act to counter scattering we have questioned if its
operation could invalidate previous conclusions concerning
subsequent terrestrial planet formation. Having updated our model to
include type I migration and having re-run three scenarios from
\citet{fogg3}, we conclude that the extra dissipation generated is
still not sufficient to prevent the buildup of a scattered disk
$\sim 50\%$ the mass of the original. Although these scattered disks
are of lower mass, they are generally less dispersed and better
damped, supporting a more rapid renewal of planetary growth. As
shown in our extended run, whilst the gas remains, residual type I
forces can act on scattered protoplanets to draw them back into the
$\sim 1$~AU region and the habitable zone. Our model also generates
planets of several Earth masses in potentially stable orbits
interior to the final orbit of the giant planet and is thus
supportive of the proposals of \citet{fogg1} and \citet{zhou} that
hot-Earths and hot-Neptunes found interior to hot-Jupiters may have
originated from material shepherded inward by a migrating giant
planet.

Hot-Jupiter systems therefore, far from being devoid of other
planetary bodies within their warmer regions, may be worthwhile
targets for future investigation.

\begin{acknowledgements}

We thank the referee, Sean Raymond, whose comments led to
improvements in this paper.

\end{acknowledgements}



\bibliographystyle{aa}

\listofobjects

\end{document}